\documentclass[sigconf,natbib=true]{acmart}
\usepackage{graphicx}
\usepackage{subfigure}
\usepackage{amsmath}
\usepackage{amsthm}
\usepackage{booktabs} 
\usepackage{enumitem}
\usepackage{multirow}
\usepackage{amsfonts}
\usepackage{wrapfig}
\usepackage{caption}

\AtBeginDocument{%
  \providecommand\BibTeX{{%
    \normalfont B\kern-0.5em{\scshape i\kern-0.25em b}\kern-0.8em\TeX}}}

\setcopyright{acmcopyright}
\copyrightyear{2018}
\acmYear{2018}
\acmDOI{XXXXXXX.XXXXXXX}

\acmConference[Conference acronym 'XX]{Make sure to enter the correct
  conference title from your rights confirmation emai}{June 03--05,
  2018}{Woodstock, NY}
\acmPrice{15.00}
\acmISBN{978-1-4503-XXXX-X/18/06}

\begin{document}

\title{Preference or Intent? \\Double Disentangled Collaborative Filtering}

\author{Chao Wang}
\affiliation{\institution{HKUST Fok Ying Tung Research Institute, The Hong Kong University of Science and Technology (Guangzhou)} \country{China}}
\email{chadwang2012@gmail.com}

\author{Hengshu Zhu}
\affiliation{%
	\institution{Career Science Lab, BOSS Zhipin}
	\country{China}
}
\email{zhuhengshu@gmail.com}

\author{Dazhong Shen}
\affiliation{%
	\institution{School of Computer Science, University of Science and Technology of China}
	\country{China}
}
\email{sdz@mail.ustc.edu.cn}

\author{Wei Wu}
\affiliation{%
	\institution{School of Computer Science, University of Science and Technology of China}
	\country{China}
}
\email{urara@mail.ustc.edu.cn}

\author{Hui Xiong}
\affiliation{%
	\institution{The Hong Kong University of Science and Technology (Guangzhou), HKUST Fok Ying Tung Research Institute}
	\country{China}}
\email{xionghui@ust.hk}


\begin{abstract}
People usually have different intents for choosing items, while their preferences under the same intent may also different. In traditional collaborative filtering approaches, both intent and preference factors are usually entangled in the modeling process, which significantly limits the robustness and interpretability of recommendation performances. For example, the low-rating items are always treated as negative feedback while they actually could provide positive information about user intent. To this end, in this paper, we propose a two-fold representation learning approach, namely Double Disentangled Collaborative Filtering (DDCF), for personalized recommendations. The first-level disentanglement is for separating the influence factors of intent and preference, while the second-level disentanglement is performed to build independent sparse preference representations under individual intent with limited computational complexity. Specifically, we employ two variational autoencoder networks, intent recognition network and preference decomposition network, to learn the intent and preference factors, respectively. In this way, the low-rating items will be treated as positive samples for modeling intents while the negative samples for modeling preferences. Finally, extensive experiments on three real-world datasets and four evaluation metrics clearly validate the effectiveness and the interpretability of DDCF.
\end{abstract}

%


\maketitle

\section{Introduction}

As one of the most popular techniques for building personalized recommender systems, collaborative filtering (CF) aims to model the user behaviors by learning latent representations based on the historical interaction records, such as the implicit feedback or explicit rating matrix. In real-world recommender systems, users usually have different intents over some item groups, while their preference on a specific item under distinct intents may also different. Marketing studies have shown that purchase intent should be distinguished from product preference and a person may indicate his preference without any intent of buying~\cite{sirgy2015self,landon1974self}. However, in traditional CF approaches, both intent and preference factors are usually entangled as integrated factors in the rating modeling process~\cite{ricci2011introduction}, which significantly limits the robustness and accuracy of recommendation results. A piece of evidence is that when transforming explicit feedback into implicit form, using all of the feedback as positive samples for training sometimes can lead to better recommendation performances than only considering high ratings as positive samples. This is counterintuitive since the low-rating interactions are always viewed as negative in the literature.

\begin{figure*}[t]
	\centering
	\includegraphics[width=0.75\textwidth]{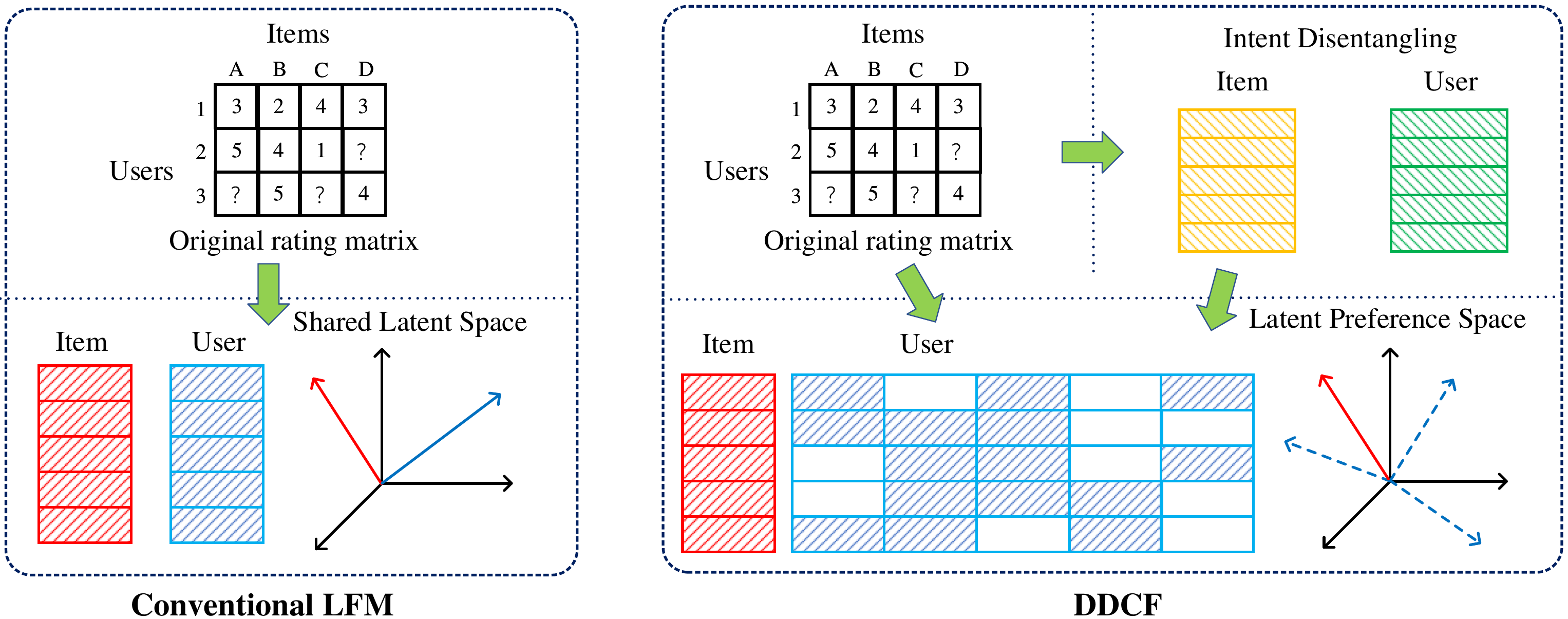}
	\vspace{-1ex}
	\caption{The rating modeling processes in conventional LFM and DDCF, respectively.}
	\label{fig:latent}
	\vspace{-2ex}
\end{figure*}

Therefore, we argue that such low-rating feedback is not completely ``negative'' since it can still provide positive guidance from the intent perspective. In this paper, we denote a user intent as one's interests on a group of items sharing similar properties, while the user preference represents her evaluation on a specific item. For instance, Alice may choose to watch a comedy \emph{A} rather than a tragic film \emph{B} for celebrating the holiday, although she actually would give a higher rating on film \emph{B} than \emph{A}.
Consequently, neglecting the user intents will lead to many limitations: 1) It is prone to inappropriately analyze the underlying preferences and thus produce suboptimal representations for both users and items; 2) Since the real-world data often include some noisy or ambiguous interaction records, the latent representations may be quite obscure and entangled to fit the integrated intents; 3) The representations are usually lack of interpretability without consideration of intents. 

However, it is very challenging to discriminate the different influence factors from massive user behaviors. First, the intents of users are implicit without tagged labels, we usually have to figure out their intents based only on the user-item rating matrix. 
In the literature, although there have been some efforts to capture the bias factors in the rating modeling process, such as popularity bias, conformity bias, selection bias and so on~\cite{timmermans2018track,papakyriakopoulos2020political,edizel2020fairecsys,saito2020unbiased,zhang2021causal,wei2021model,wan2022cross},
the task of disentangling the intent and preference is still under-explored. Recently, ~\citet{ma2019learning,wang2020disentangled} began to study users' intents over item groups in implicit feedback. However, they did not clearly disentangle the intent and preference factors, and can only handle quite coarse-grained intents due to the unacceptable computational complexity with respect to the number of intents. Besides, the disentanglement of user preference representations under different intents is also vital for producing high-quality representations, which has still been largely neglected in the literature.

To this end, in this paper, we propose a novel CF approach for personalized recommendation, namely Double Disentangled Collaborative Filtering (DDCF), which can perform two-fold disentanglement in the rating modeling process for explicit feedback and thus produce more robust and interpretable latent representations. Formally, in this paper, we define one intent channel as a probability distribution over all items, and one user’s intent distribution is a probability distribution over all intent channels. Hence, the items with high probabilities in one intent channel can be viewed as a group. Accordingly, our model separates the traditional integrated latent space into two new spaces, representing the intent and preference factors, respectively. Each of both is learned from a variational autoencoder (VAE) style network, namely the intent recognition network and preference decomposition network. We first learn the distributions over all the intent channels for both users and items from the implicit-form input. Then, the observed user ratings will be partitioned into each intent based on the learned user and item intent distributions. Thus, we can obtain different latent preference representations from those tailored ratings for different concepts by the preference decomposition network. For handling fine-grained intent channels, we propose to adopt the independent sparse representations for user preferences through sampling technology so that the computational complexity would not grow with the number of intent channels. Besides, we designed a novel disentangled contrastive Learning mechanism for learning preference representations. Meanwhile, we also discuss some new recommendation strategies based on the intent representations. To summarize, the main contributions of this work are:

\begin{itemize}
	\setlength{\itemsep}{0pt}
	\setlength{\parsep}{0pt}
	\setlength{\parskip}{0pt}
	\item We discuss and emphasize that low-rating feedback is not completely ``negative''. In our approach, the low-rating items will be treated as positive samples for modeling intents while negative samples for modeling preferences.
	\item We propose a novel double disentangled collaborative filtering approach, DDCF. Specifically, the first-level disentanglement is designed for separating the influence factors of intent and preference, while the second-level disentanglement is performed to construct independent sparse representations for different intents.
	\item We propose a novel disentangled contrastive Learning mechanism to improve the disentangled embedding quality of user preference under different intent channels.
	\item We validate the effectiveness and interpretability of our approach DDCF on three real-world datasets and a number of state-of-the-art baselines.
	
\end{itemize}

\vspace{-2ex}
\section{Preliminary and Related Works}\label{neg}
In this section, we first discuss and emphasize the limitations of the representation learning process in conventional latent factor models (LFMs) with respect to the modeling of the user-item rating matrix. Then, we will introduce some related works about the improvements of the representation learning process in LFMs.

\vspace{-2ex}
\subsection{Rating Modeling Paradigm}
Collaborative filtering (CF)~\cite{mnih2008probabilistic,melville2010recommender} methods have been widely applied in many web-based services like Google, Netflix, and Amazon~\cite{dror2011yahoo}. Among them, latent factor model (LFM) based CF methods have drawn many researchers' attention and achieved great success in recommender systems due to their superior recommendation performance and high extensibility~\cite{rao2015collaborative,barkan2016item2vec,wang2016collaborative}. Generally speaking, the user-item rating matrix can be considered as a low-rank matrix with a vast majority of unobserved entries. Supposing there are $ N $ users and $ M $ items in the data. The main idea of the latent factor model is to factorize the low-rank rating matrix $R \in \mathbb{R}^{N\times M}$ into two latent representations $U \in \mathbb{R}^{d\times N}$ and $V \in \mathbb{R}^{d\times M}$ in a shared low-dimensional space with dimension $d$, representing the latent user and item factors, respectively. Further, we respectively use $u_i\in \mathbb{R}^{d}$ and $v_j\in \mathbb{R}^{d}$ to denote the latent vectors for the $ i $-th user and $ j $-th item. As shown in Figure~\ref{fig:latent}, $ u_i $ implies the $ i $-th user's preferences while $ v_j $ represents the $ j $-th item's properties. When $ u_i $ and $ v_j $ get closer in the latent space, the $ i $-th user would be more likely to prefer the $ j $-th item. Let $ f_{e}(\cdot) $ denote the embedding function and $ i $, $ j $ be the user and item's IDs. Here, we summarize the representation learning process of conventional LFMs as follows:
\begin{eqnarray}
	&u_i = f_{e}(i),
	&v_j = f_{e}(j).
\end{eqnarray}

After obtaining the latent representations of both users and items, the next step is to reconstruct the user-item interactions through the rating modeling process. The most widely-used rating prediction paradigm is the inner product of latent user vector and item vector\cite{koren2009matrix,xue2017deep}:
\begin{equation}\label{equ:rating}
	\hat{R}_{ij} = u_i^T v_j,
\end{equation}
where $ \hat{R}_{ij} $ denotes the predicted rating for the $ i $-th user on the $ j $-th item. Equation~\ref{equ:rating} implies that we measure the user-item preferences by the distance estimation between the two latent vectors in the shared latent space as shown in Figure~\ref{fig:latent}.

\begin{table}[!t]
	\caption{Performances with/without negative items. }
	\vspace{-2ex}
	\label{tab:com}
	\centering
	\resizebox{7.8cm}{!}{
		\begin{tabular}{llcccccc}
			\toprule
			Datasets & Methods  & P@10  & R@10 & MAP@10 \\
			\midrule
			\multirow{6}{0.5cm}{\emph{MovieLens}} & REL (Setting 1)   & $ 0.4016 $  & $ {0.1018} $ & $ 0.2922 $      \\
			&REL (Setting 2) 	& $ 0.3829 $ &  $ {0.0906} $ & $ 0.2760 $    \\
			& PDA (Setting 1)   & $ 0.4064 $  & $ {0.1012} $ & $ 0.2965 $      \\
			&PDA (Setting 2) 	& $ 0.3916 $ &  $ {0.0936} $ & $ 0.2836 $    \\
			& DMF (Setting 1)   & $ 0.4302 $  & $ {0.1205} $ & $ 0.3112 $      \\
			& DMF (Setting 2) 	& $ 0.4282 $ &  $ {0.1104} $ & $ 0.3109 $      \\
			&Multi-VAE (Setting 1) & $ 0.4596 $ & $ {0.1232} $ & $ 0.3504 $   \\
			&Multi-VAE (Setting 2)  & $ 0.4533 $ & $ {0.1126} $ & $ 0.3494 $      \\
			\midrule
			\multirow{6}{0.5cm}{\emph{Amovie}} & REL (Setting 1)   & $ 0.0532 $  & $ {0.0433} $ & $ 0.0251 $      \\
			&REL (Setting 2) 	& $ 0.0444 $ &  $ {0.0353} $ & $ 0.0205 $    \\
			& PDA (Setting 1)   & $ 0.0588 $  & $ {0.0475} $ & $ 0.0290 $      \\
			&PDA (Setting 2) 	& $ 0.0505 $ &  $ {0.0391} $ & $ 0.0242 $    \\
			& DMF (Setting 1)  & $ 0.0733 $ & $ {0.0593} $ & $ 0.0532 $      \\
			&DMF (Setting 2) 	& $ 0.0677 $  & $ {0.0545} $ & $ 0.0342 $ \\
			&Multi-VAE (Setting 1) & $ 0.0765 $ & $ {0.0643} $ & $ 0.0396 $    \\
			&Multi-VAE (Setting 2) & $ 0.0714 $  & $ {0.0584} $ & $ 0.0365 $   \\
			\midrule
			\multirow{6}{0.5cm}{\emph{Yahoo}} & REL (Setting 1)   & $ 0.1070 $  & $ {0.3106} $ & $ 0.0536 $      \\
			&REL (Setting 2) 	& $ 0.0998 $ &  $ {0.2789} $ & $ 0.0497 $    \\
			& PDA (Setting 1)   & $ 0.1038 $  & $ {0.3168} $ & $ 0.0567 $      \\
			&PDA (Setting 2) 	& $ 0.0970 $ &  $ {0.2869} $ & $ 0.0514 $    \\
			& DMF (Setting 1)  & $ 0.1113 $ & $ {0.3321} $ & $ 0.0592 $     \\
			&DMF (Setting 2) 	& $ 0.1075 $ & $ {0.3050} $ & $ 0.0567 $     \\
			&Multi-VAE (Setting 1) & $ 0.1140 $ & $ {0.3369} $ & $ 0.0622 $ \\
			&Multi-VAE (Setting 2)   & $ 0.1092 $ & $ {0.3121} $ & $ 0.0590 $  \\
			\bottomrule
	\end{tabular}}
\end{table}

However, Equation~\ref{equ:rating} failed to consider and disentangle the intent factor in user behaviors. Thus, the learned latent representations will lack interpretability and may miss some important information.

\subsection{Entangled Data Analysis}
In this subsection, we leveraged four representative CF methods for implicit feedback, REL~\cite{saito2020unbiased}, PDA~\cite{zhang2021causal}, DMF~\cite{xue2017deep}, and Multi-VAE~\cite{liang2018variational}, to perform an interesting comparison between two settings. 
Since many public real-world recommendation data were provided in the form of explicit feedback, researchers usually chose to transform the graded ratings into binary data for studying implicit feedback\cite{hu2008collaborative,rendle2009bpr,he2017neural,wu2018sql}. Accordingly, we have to distinguish the positive and negative feedback from original ratings. In the literature, the most common treatment method is to set a threshold\cite{hu2008collaborative,rendle2009bpr,he2017neural,wu2018sql}. In this way, the items with ratings lower than the given threshold would be viewed as negative feedback. However, an important question raises, ``Should the items with low ratings be viewed as totally negative samples?''

In our experiment, for the first setting, we transformed all the explicit feedback into positive feedback in the training sets, while only the high ratings were transformed into positive feedback for the second setting. For the test set, only high ratings were viewed as positive feedback for both settings. Thus, the two settings share the same test set and different training sets in this experiment. To perform a fair comparison, we fix all the hyper-parameters to be the same in the two settings. We present the experimental results in the two settings on the three datasets in Table~\ref{tab:com}.
Surprisingly, we can observe from Table~\ref{tab:com} that the performances in the first setting are better than in the second setting. These results clearly demonstrate the conventional ’negative feedback’ (items with low ratings) are actually not totally negative. Actually, these interactions also provide the intent information behind users’ decision making processes. Without the information of low-rating items, we cannot make a full understanding of all the user's intents, and thus the CF models would perform worse on covering user interests. A piece of evidence is that the difference on the recall metric is larger than the precision metric.

It is worth noting that the user intent are quite different from the well-known data bias problem in recommender systems. The intents comes from personalized user inherent interests, while data bias represents the distribution for which the training data is collected is different from the ideal test data distribution~\cite{chen2020bias}. To avoid the influence of data bias, we leveraged two debiasing approaches, REL and PDA. REL is for Missing-Not-At-Random problem and PDA is for conformity/popularity bias. These biases may lead to the improvements of performance with more negative samples. However, we can observe from Table~\ref{tab:com} that these two debiasing approaches have significant differences between the two settings, demonstrating the differences do not come from data bias problem.

To summarize, while a low rating may imply low user preference over the item itself, it would also indicate potential user intent, i.e., the interest over the groups the item belongs to. Therefore, our motivation is to disentangle the intent and preference in the user-decision process. 

\vspace{-2ex}
\subsection{Related Works}
Since exploring the user choices based purely on the rating matrix is the central task of LFM, there are many efforts trying to improve the rating modeling process in the literature.
some researchers pointed out that the straightforward rating modeling process is insufficient for depicting complex real-world situations. For example, \citet{wang2018modeling} and \citet{wang2018collaborative} believed that the exposure status in real-world situations would significantly influence the user-item interaction probabilities. Thus, they analyzed the exposure problem for missing data by estimating user exposures towards items, and then used them to guide the rating modeling process. Besides, \citet{abdollahpouri2017controlling} and \citet{yang2018unbiased} discussed the popularity factor in real-world data that users are more likely to interact with popular items. There are also many other works focused on different biases in the rating modeling process\cite{burke2018balanced,timmermans2018track,hofmann2014effects,papakyriakopoulos2020political,edizel2020fairecsys,saito2020unbiased,zhang2021causal,wei2021model,wan2022cross}.

Although there have been many efforts in the rating modeling process, the disentanglement of intent and preference still largely remains unexplored, which restricts the further understanding of the latent representations. Recently, \citet{ma2019learning} and \citet{wang2020disentangled}attempted to construct independent representation for each intent in implicit feedback. \citet{li2021intention} aimed to mine the latent user intents in sequential data. \citet{li2021intention} and \citet{zhao2022multi} leverage graph modeling to study the user intents behind user feedback. Differently, \citet{wang2021learning} tried to model intents with knowledge graph but not interactions. However, these methods cannot explicitly learn the disentangled intent and preference representations. They are mostly designed for implicit feedback and incapable of differentiating the influence of intent and preference. Besides, they are mainly suitable for handling coarse-grained intents due to the at least linear computational complexity to the number of intents. 

In this paper, 
in order to obtain more robust and interpretable user and item representations, we focus on the two-folded disentanglement implementation for explicit feedback and innovatively propose a double-disentangled collaborative filtering approach to concurrently consider both the users' intents and preferences.

\vspace{-2ex}
\section{Methodology}
In this section, we will introduce the technical details of our proposed Double Disentangled Collaborative Filtering (DDCF) approach. We will begin with the notations and solution overview.

\vspace{-2ex}
\subsection{Solution Overview}
An explicit rating matrix $ R \in \mathbb{R}^{N\times M} $ is composed of historical ratings with real numbers and other missing entries denoted as 0.
Since the ratings may have various ranges in real-world applications, here we suppose all the ratings are larger than 0 without loss of generality.
Let $ R_{i} $ denote the $ i $-th user's rating records on all items. Then we use the binary matrix $ X \in \mathbb{R}^{N\times M} $ to denote the implicit form of $ R $, that is, $ X_{ij} = 1 $ if $ R_{ij} > 0 $ and $ X_{ij} = 0$ if $R_{ij} = 0 $. Thus, all the user-item interactions are treated equally in the binary matrix $ X $ while the rating matrix $ R $ contains more information about users' individual preferences. Similarly, the $ i $-th user's binary records can be represented by $ X_{i} $.

As shown in Figure~\ref{fig:framework}, DDCF aims to perform double disentangled representation learning for better modeling the user-item interactions. Along this line, we learn two types of latent representations for both users and items, namely intent representation and preference representation. Specifically, we first adopt input $ X $ in the intent recognition network to model the intent representation, which represents the user and item's distribution on each intent channel. Formally, we define one intent channel as a probability distribution over the items. Then, based on the obtained intent vectors, we further decompose the user's rating record $ R_{i} $ into tailored inputs $ R_{il}\in \mathbb{R}^{M} $ according to different intent channels. To solve the complexity problem, we use the sampling technique and only learn the independent preference representations $ u_{il}\in \mathbb{R}^{d} $ for top-$ L $ intent channels through the preference decomposition network. To better mine the user preference under different intention channels, we design the disentangled contrastive learning mechanism, which share the same network parameter $ \Theta $ with preference decomposition network. Finally, the rating prediction is performed by jointly considering the predicted ratings under $ L $ intent channels.

\vspace{-2ex}
\subsection{Intent Recognition Modeling}
The user behavior may be driven by multiple intent channels. Taking movie recommendations as an example, the appeal to users may come from movies' categories, directors, actors, plots, and many other fine-grained concepts. Hence, distinct intent channels will lead to distinct contributions to the final user decisions.

The first step of DDCF is to model the users' and items' distributions on all the intent channels. Suppose there are $ K $ intent channels in the data. Inspired by classic topic modeling algorithm~\cite{blei2003latent}, we assume every user binary records $ X_i $ is generated from a mixture of $ K $ intent channels $ \beta = (\beta_1,...,\beta_K) \in \mathbb{R}^{M\times K}$. Each intent channel $ \beta_k \in \mathbb{R}^{M}$ can be viewed as a probability distribution over the entire $ M $ items. A higher probability value in $ \beta_k $ implies this item is more likely to belong to this channel, where the sum of all the items' probability values should be equal to 1 for each channel.
Then the user intent representation $ \gamma_i \in \mathbb{R}^{K} $ for the $ i $-th user is defined as the proportion distribution over all the channels, which obeys Dirichlet distribution: $ \gamma_i \sim Dirichlet(\alpha)$. Here $ \alpha \in \mathbb{R}^{K} $ is the learned parameter and the $ k $-th dimension of $ \gamma_i $ represents the proportion of the $ k $-th intent channel.

\begin{figure*}[t!]
	\centering
	\includegraphics[width=0.88\textwidth]{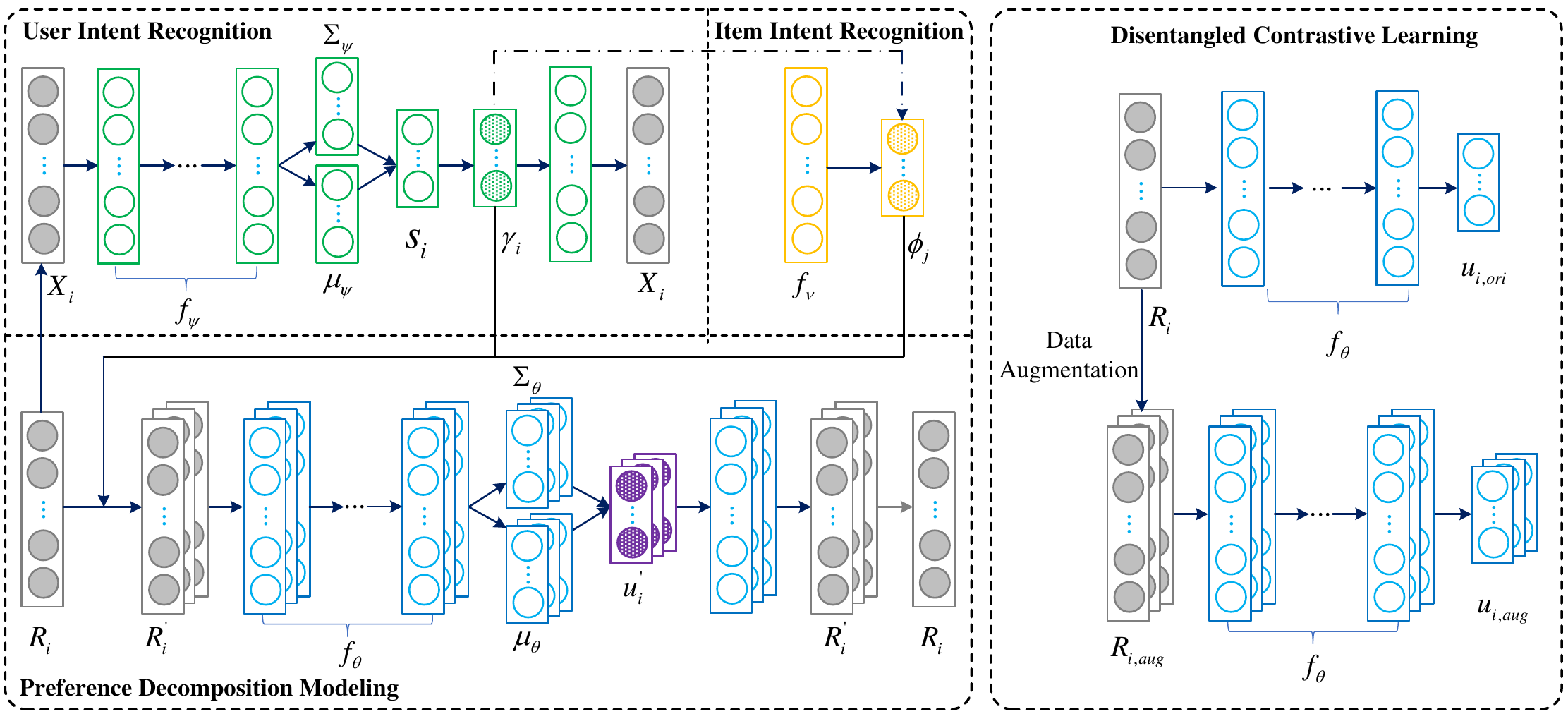}
	\vspace{-2ex}
	\caption{The network architectures of DDCF.}
	\label{fig:framework}
	\vspace{-2ex}
\end{figure*}

Supposing there are $ N_i $ positive items in the user binary records $ X_i $.
Hence, $ X_{i} $ can be drawn from the Multinomial distribution: $ X_{i} \sim Multinomial(N_i, \beta\gamma_i) $. Under this assumption, the marginal likelihood of the user binary record $ X_i $ is given as follows:

\vspace{-2ex}
\begin{equation}\label{equ:intent}
	p(X_i|\alpha, \beta)=\int_{\gamma_i}p(X_{i}|\beta,\gamma_i)p(\gamma_i|\alpha)d\gamma_i.
\end{equation}
\vspace{-2ex}

Note that the Dirichlet prior is significant for obtaining interpretable and sparse intent representations~\cite{wallach2009rethinking}. However, it is incapable of directly taking gradients through the sampling process.
Therefore, we utilize a Laplace approximation to the softmax basis of Dirichlet prior~\cite{mackay1998choice} and then perform the reparameterization trick for the Normal distribution.
Specifically, let $ \gamma_i = \sigma(s_i/\tau) $, where $ \sigma(\cdot) $ is the softmax function with temperature parameter $ \tau $. Each dimension of $ s_i $ is related to a specific intent channel in full accord with $ \gamma_i $. Then $ s_i $ can be given as a multivariate Normal with mean $ \mu $ and covariance $ \Sigma $. Note that the covariance matrix is an approximately diagonal covariance matrix for large $ K $ since its off-diagonal elements will be suppressed with $ O(1/K) $. Hence, the Laplace approximation $ p(s_i) $ can be given by:

\vspace{-2ex}
\begin{small}
	\begin{eqnarray}\label{equ:laplace} \mu_k=\log\alpha_k-\frac{1}{K}\sum_{k'=1}^K \log\alpha_{k'}, \quad
		\Sigma_{kk}=\frac{1}{\alpha_k}(1-\frac{2}{K})+\frac{1}{K^2}\sum_{k'=1}^K\frac{1}{\alpha_{k'}}.
	\end{eqnarray}
\end{small}
\vspace{-2ex}

Since the latent vector $ s_i $ is the softmax basis of Dirichlet prior, we are able to approximate the simplex basis by the logistic normal distribution $ p(\gamma_i|\alpha) \approx \mathcal{LN}(\gamma_i|\mu, \Sigma) $~\cite{lafferty2006correlated}.	

According to Equation~\ref{equ:intent}, the generation network is defined as a linear layer $ \hat{X_i} = \beta\gamma_i $. Here $ \beta $ should be constrained to make sure the sum of each column is equal to 1 in principle, which can be easily realized by applying the softmax transformation to $ \beta $. 

To inference the model, we employ the Neural Variational inference (NVI) approach due to its success in the embedding competency~\cite{kingma2013auto,srivastava2017autoencoding}.
Specifically, we construct an inference network
parameterized by $ \psi $ for efficient posterior inference of $ s_i $.

\vspace{-2ex}
\begin{equation}
	q(S) = \prod_i q_\psi(s_i|X_i)=\prod_i \mathcal{N}(\mu_\psi(X_i),\Sigma_\psi(X_i)).
\end{equation}
\vspace{-2ex}

Here the variational distribution obeys multivariate Normal distribution owing to the Laplace approximation. The prior distribution $ p(s_i) $ is also given in the form of Equation~\ref{equ:laplace}, where all the $ \alpha_k $ is set as a predefined value. Then, we perform the inference network $ f_\psi(\cdot) $ through an $ E $-layer MLP. As shown in Figure~\ref{fig:framework}, the network input is $ X_i $. The first layer can be seen as the embedding layer since the input is in the $ 0/1 $ form. Each layer contains the linear transformation and activation function except for the last layer, which produces two outputs, $ \mu_\psi(X_i) $ and $ \Sigma_\psi(X_i) $. Accordingly, by leveraging the reparameterization trick~\cite{rezende2014stochastic}, we can easily obtain $ q(s_i) $ by $ s_i = \mu_\psi(X_i) + \epsilon \Sigma_\psi(X_i)$, where $ \epsilon \sim \mathcal{N}(0,I) $. Finally, we can write the evidence lower bound (ELBO) as

\vspace{-2ex}
\begin{align}\label{equ:elbo}
	-\mathcal{L}_1
	=\mathbb{E}_q[\log p_\psi(X|S)] - \eta\mathbb{KL}(q_\phi(S|X)||p(S)).
\end{align}
\vspace{-2ex}

The former term in Equation~\ref{equ:elbo} is the reconstruction loss. Following VAE~\cite{kingma2013auto}, we employ the Monte Carlo sampling to estimating the expected values $ E_q[\log p(X|S)]= \sum_i\frac{1}{H}\sum_h p_\psi(X_i|s_i^{(h)}) $. The latter term in Equation~\ref{equ:elbo} is the KL divergence, which can be given in analytical form: 
$\nonumber \mathbb{KL}(q_\psi||p) = \frac{1}{2}\sum_i [(\mu-\mu_\psi(X_i))^T\Sigma^{-1}(\mu-\mu_\psi(X_i))
+ tr(\Sigma^{-1}\Sigma_\psi(X_i)) + \log\frac{|\Sigma|}{|\Sigma_\psi(X_i)|}-K].$ Furthermore, to encourage the independence among different dimensions in each representation, we follow $ \beta $-VAE~\cite{higgins2016beta} to add a penalty parameter $ \eta $ for the KL term. Such treatment could help the representation ignore noise in the input and focus more on the key information~\cite{burgess2018understanding}.

As for obtaining the item intent representation $ \phi\in \mathbb{R}^{K\times M} $, naturally we can assume the user representation $ \gamma_i $ as the prior probability for the interacted items, i.e., $ j\in \{j|X_{ij}=1\} $. Then we employ the MLP $  f_\nu(W_j)$ as the posterior inference network:

\vspace{-2ex}
\begin{align} \phi_j \sim Multinomial(1, \sigma(f_\nu(W_j))/\tau). 
\end{align}
\vspace{-2ex}

Here $ W_j $ is the $ j $-th row of the weight matrix in the first layer of the network $ f_\psi $ so that the initial embeddings of the items are shared in the two inference networks. Consequently, we use the KL divergence to measure the distance between the prior and posterior probability as follows:

\vspace{-2ex}
\begin{align}\label{equ:l2}
	\mathcal{L}_2
	=\Sigma_i\Sigma_{j\in\{j|X_{ij}=1\}} \sigma(f_\nu(W_j)/\tau)^T \log \frac{\sigma(f_\nu(W_j)/\tau)}{\gamma_i}.
\end{align}
\vspace{-2ex}

\subsection{Preference Decomposition Modeling}
In this subsection, for constructing disentangled preference representations, we will further decompose the user preferences into the corresponding intent channels.

The user-item rating $ R_{ij} $ can be viewed as the composite outcome of the $ i $-th user's preference under different intent channels. Therefore, we first decompose the user ratings $ R_i $ into each channel. Specifically, the input of the preference decomposition network $ R_{il} $ is defined as $ R_{il} = L2norm(\phi^T_l \cdot R_i) $, where $ \cdot $ represents the inner product and $ \phi^T_l \in \mathbb{R}^{ M}$ is the $ l $-th row of matrix $ \phi $. We also perform L2 normalization to rescale the inputs. In this way, the modified rating order among the items may be quite different from the observed order. For example, item A is rated with 5 points while item B is rated with 3 points. However, under the channel which is quite related to item B while not related to item A, the modified rating may be 0.4 and 0.7 for item A and B, respectively.

Different from traditional LFMs, the disentangled preference representation $ u_i \in \mathbb{R}^{d'}, d'=Kd$ is composed of the independent representations $ u_{il}\in \mathbb{R}^{d} $ as shown in Figure~\ref{fig:latent}. Specifically, each subdivided representation $ u_{il} $ means the preference under the $ l $-th intent channel in the latent preference space. However, when the number of channels $ K $ becomes too large, the total dimension of $ u_i $ will be unacceptable. To avoid this problem, considering that the top channels have the biggest influences for users’ decision making processes, we first obtained the sampled intent distribution $ \gamma_i$ for the $ i $-th user. Then we only pick the top-$ L $ channels from $ \gamma_i $ for constructing preference representations, where $ L $ is a very small number. Along this line, the obtained user representation matrix $ u_i $ will be a sparse matrix, since only $ Ld $ entries in each row are non-zero as shown in Figure~\ref{fig:latent}, where $ L << K $. Note that the selected $ L $ intent channels will change with different sampling results to avoid falling into local optimum.

For convenience, we let $ R'_i \in\mathbb{R}^{L\times M}$ and $ u'_i \in\mathbb{R}^{L\times d}$ denote the processed input and user representation for the $ i $-th user with $ L $ intent channels. Now we can detailedly introduce the preference decomposition network, which is also based on VAE architecture as shown in Figure~\ref{fig:framework}. Firstly, the encoder $ f_\theta(\cdot) $ transforms the $ L $ modified inputs $ R_{il} $ to produce independent representations $ u_{il} $ for each user. Here we also select MLP as the network architecture. Each layer consists of the linear transformation and activation function except for the last layer. The last layer generates two outputs, $ \mu_\theta(R'_i) \in\mathbb{R}^{L\times d} $ and $ \Sigma_\theta(R'_i) \in\mathbb{R}^{L\times d\times d} $. The prior $ p_\theta(u_{il}) $ was chosen as the standard normal distribution $ \mathcal{N}(\mathbf{0},\mathbf{I}) $. Accordingly, $ u'_i $ can be drawn by the variational distribution as follows:

\vspace{-2ex}
\begin{equation}
	q_\theta(U') = \prod_i q_\theta(u'_i|R'_i)=\prod_i \prod_l \mathcal{N}(\mu_\theta(R_{il}),\Sigma_\theta(R_{il})).
\end{equation}
\vspace{-2ex}

The decoder aims to restore the input $ R_{i} $ by the corresponding processed representation $ u'_{i} $. Following the LFM, we let $v_j\in \mathbb{R}^{d}$ denote the latent property presentation for the $ j $-th item. Hence, the decoder is defined by the inner product of the user and item representations, $ \hat{R}_{ijl} = u_{il}^Tv_j $. To gather the predicted ratings under different intent channels, a simple but effective manner is the weighted average. Specifically, we first choose the top-$ L $ largest dimensions in $ \gamma_i $ to form up the processed weight vector $ \gamma_i' $. Then we perform the normalization treatment to make sure that the sum of all the dimensions of $ \gamma_i' $ is equal to 1. Consequently, we have 

\vspace{-2ex}
\begin{align}
	\hat{R}_{ij} = \Sigma_l \frac{\gamma'_{il}}{\Sigma_{l'}\gamma'_{il'}}\hat{R}_{ijl} .
\end{align}
\vspace{-2ex}

Note that all the weights and biases in both encoder and decoder share the same parameters among different intent channels so that the number of entire trainable parameters is equal to a common autoencoder network. Also, the computational complexity of the entire network is limited since $ L $ is very small.

Finally, we can give the ELBO of the preference decomposition network as follows:

\vspace{-2ex}
\begin{align}\label{equ:l3}
	-\mathcal{L}_3
	=\mathbb{E}_q[\log p_\theta(R'|U')] - \eta\  \mathbb{KL}(q_\theta(U'|R')||p(U')).
\end{align}
\vspace{-2ex}

\vspace{-2ex}
\subsection{Disentangled Contrastive Learning}\label{contra}
To better study the consistency between the original feedback and different views, in this subsection, we design a disentangled contrastive learning mechanism. Firstly, we leverage the encoder $ f_\theta(\cdot) $ to transform the original input $ R_i $ into user embedding $ u_{i,ori} $. Then, following~\citet{wu2021self}, we apply data augmentation tricks, node and edge dropout, to $ L $ modified inputs $ R_{il} $ to produce augmented inputs $ R_{il,aug} $ for each user. Again, we leverage the encoder $ f_\theta(\cdot) $ to transform $ R_{il,aug} $ into disentangled user representation $ u_{il,aug} $. Note that the network $ f_\theta(\cdot) $ share the same parameters with the encoder in preference decomposition modeling network. 

The augmented user input $ R_{il,aug} $ can be viewed as different views of the original user feedback $ R_i $ under different intent channels. 
Conventional contrastive learning methods, such as InfoNCE~\cite{gutmann2010noise}, usually treat two views generated from the same sample as positive pairs. However, in DDCF, the user representations under different intent channels should be distinct. Instead, we treat $ u_{i,ori} $ and each of the $ L $ disentangled representations $ u_{il,aug} $ as a positive pair. $ u_i,ori $ and the other users' representations $ u_{i'l,aug}, i' \neq i $ are treated as negative pairs, enforcing the divergence among different users. Formally, we can maximize the agreement of positive pairs and minimize that of negative pairs with the following contrastive loss:

\vspace{-2ex}
\begin{align}\label{equ:l4}
	\mathcal{L}_4
	=\Sigma_i \Sigma_l -\log \frac{\exp(\cos(u_{i,ori}, u_{il,aug})/\tau_{c})}{\Sigma_{i' \neq i} \exp(\cos(u_{i,ori}, u_{i'l,aug})/\tau_{c})}.
\end{align}
\vspace{-2ex}

Here the cosine function $ \cos(\cdot) $ measures the similarity between two vectors. $ \tau_{c} $ is the temperature parameter.

\vspace{-2ex}
\subsection{Optimization}\label{optim}
In this subsection, we will detailedly introduce how to optimize the four losses given above. The optimization process can be divided into two stages, pre-training stage and unified learning stage.

Firstly, since the preference decomposition modeling process is based on the intent recognition modeling process, the learned disentangled preference representation $ u_i $ would be meaningless if the intent distribution $ \gamma_i $ is randomized. Therefore, in order to avoid converging to the local optimum prematurely, we need to first pre-train the intent recognition networks with the two loss functions $\mathcal{L}_1  $ and $ \mathcal{L}_2 $.

Note that in Equation~\ref{equ:l2}, we leverage the KL divergence between each user's intent representation and interacted items' intent representations. Hence, the loss $ \mathcal{L}_2 $ has a trend to let all the intent channels have similar probabilities
, which obviously run in opposite directions against our motivation of disentanglement. Here we introduce the stop-gradient strategy to prevent this problem. Specifically, we can stop the gradient of $ \gamma_i $ in Equation~\ref{equ:l2} in the backpropagation process, since we mainly aim to learn the item intent representations $ \phi_j $ with the help of user intent representations $ \gamma_i $. Let $ sg(\cdot) $ denote the stop-gradient treatment. We have:

\vspace{-2ex}
\begin{align}\label{equ:l22}
	\hat{\mathcal{L}}_2
	=\Sigma_i\Sigma_{j\in\{j|X_{ij}=1\}} \sigma(f_\nu(W_j)/\tau)^T \log \frac{\sigma(f_\nu(W_j)/\tau)}{sg(\gamma_i)}.
\end{align}
\vspace{-2ex}

In this way, the pre-training loss function can be given by:

\vspace{-2ex}
\begin{align}\label{equ:pre}
	\mathcal{L}_{pre} =\mathcal{L}_{1} + \lambda_2\hat{\mathcal{L}}_2.
\end{align}
\vspace{-2ex}

After the pre-training, we can jointly optimize the intent recognition and preference decomposition modeling processes.
Finally, the unified loss of our DDCF model is given by:

\vspace{-2ex}
\begin{equation}
	\mathcal{L} = \mathcal{L}_1+\lambda_2 \hat{\mathcal{L}}_2+\lambda_3\mathcal{L}_3+\lambda_4\mathcal{L}_4.
\end{equation}
\vspace{-2ex}

\vspace{-2ex}
\subsection{Recommendation}
The disentangled representations can promote some novel and flexible recommendation strategies to satisfy the specific needs of users when recommending in practical scenes. 
First, as in the training process, we can integrate users' intents by employing the weighted average with users' intent distributions. In this way, the obtained recommendation results come from comprehensive user interests. Besides,  we can only rank the movies under one specific intent channel or assign a user-determined intent distribution to replace the predicted user distribution for satisfying users' dynamic intents. Then we can simply rank these movies by the predicted ratings or design a new scoring system considering both similarities and preferences. Moreover, when a user wants to find some movies similar to one specific movie, we can calculate the similarity between the other movies' intent distributions and this movie's distribution to find similar movies.

\section{Experiments}
In this section, we first provide detailed information on the three datasets and evaluation protocols used in the experiments~\footnote{All the code will be publicly available after the paper is accepted.}. Next, we introduce the baseline methods and our experimental settings. Then, we report the recommendation performance results of our proposed DDCF models compared to the state-of-the-art baselines. We also discuss the influence of contrastive learning, hyper-parameters and validate the intents. Finally, we present some case studies to show the interpretability of DDCF.

\vspace{-2ex}
\subsection{Experimental Settings}
\textbf{Datasets.} We conducted our experiments on three real-world datasets, i.e., \emph{MovieLens}~\footnote{https://grouplens.org/datasets/movielens/},  \emph{Amovie}~\footnote{http://jmcauley.ucsd.edu/data/amazon/} and \emph{Yahoo}~\footnote{https://webscope.sandbox.yahoo.com/catalog.php?datatype=r}. \emph{MovieLens-20m} is a widely-used movie recommendation dataset. \emph{Amovie} is a dataset consisting of product ratings collected from Amazon Movies and TV. \emph{Yahoo}~\citep{marlin2009collaborative} contains the ratings for songs from Yahoo! Music. All the rating data in \emph{MovieLens}, \emph{Amovie}, and \emph{Yahoo} are in the form of 5-stars. For validation, following \citet{wu2018sql}, we adopted the data preprocessing to differentiate the positive and negative feedback depending on whether the ratings are not less than 4. Besides, in order to make sure we have adequate observed feedback for better evaluating the recommendation algorithms, we filtered out users with less than 10 observed items. After the data preprocessing, \emph{MovieLens} contains 20,623 users and 12,975 items with 2,980,083 observed entries. \emph{Amovie} contains 11,838 users and 9,107 items with 166,363 observed entries. \emph{Yahoo} contains 13,847 users and 1000 items with 350,174 observed entries.

\textbf{Evaluation metrics.} To construct the training set, we randomly sampled $ 60\% $ observed items for each user. Then, we sampled 10\% observed items of each user for validation, and the rest data were used for the test. Hence, we randomly split each dataset five times and reported all the results by average values. We employed four widely used evaluation metrics for evaluating the performance, i.e., P@$ K $, R@$ K $, MAP@$ K $, and NDCG@$ K $~\citep{wu2018sql}. For each user, P~(Precision)~@$ K $ measures the ratio of correct prediction results among top-$ K $ items to $ P $ and R~(Recall)~@$ K $ measures the ratio of correct prediction results among top-$ K $ items to all positive items. Furthermore, MAP (Mean Average Precision) @$ K $ and  NDCG (Normalized Discounted Cumulative Gain)~@$ K $ consider the ranking of correct prediction results among top-$ K $ items.
The results of the four metrics are given in the average of all users.

\begin{table*}[bt]
	\caption{The overall recommendation performances of different approaches. }
	\vspace{-2ex}
	\label{tab2}
	\centering
	\resizebox{14.3cm}{!}{
		\begin{tabular}{llcccccccc}
			\toprule
			Datasets & Methods & P@5 & P@10 & R@5 & R@10 & MAP@5 & MAP@10 & NDCG@5 & NDCG@10\\
			\midrule
			\multirow{10}{2cm}{\emph{MovieLens}} & PMF  & $ 0.1863 $ & $ 0.1650 $ & $ 0.0336 $ & $ {0.0558} $ & $ 0.1318 $ & $ 0.0995 $ & $ 0.1955 $ & $ 0.1843 $     \\
			&Primal-CR++	& $ 0.2872 $ & $ 0.2587 $ & $ 0.0582 $ & $ {0.0934} $ & $ 0.2250 $ & $ 0.1804 $   & $ {0.3035} $ & $ {0.2907}$    \\
			&DGCF  & $ 0.3318 $ & $ 0.2991 $ & $ 0.0722 $ & $ {0.1211} $ & $ 0.2591 $ & $ 0.2081 $  & $ {0.3487} $ & $ {0.3371} $    \\
			&SQL-Rank & $ 0.3405 $ & $ 0.3034 $ & $ 0.0750 $ & $ {0.1224} $ & $ 0.2678 $ & $ 0.2140 $  & $ {0.3585} $ & $ {0.3437} $    \\
			&MACR  & $ 0.3571 $ & $ 0.3164 $ & $ 0.0851 $ & $ {0.1403} $ & $ 0.2840 $ & $ 0.2257 $ & $ {0.3895} $ & $ {0.3703}$      \\
			&CPR  & $ 0.3437 $ & $ 0.3087 $ & $ 0.0909 $ & $ {0.1510} $ & $ 0.2635 $ & $ 0.2097 $ & $ {0.3825} $ & $ {0.3711} $     \\
			&DMF  & $ 0.4033 $ & $ 0.3589 $ & $ 0.1019 $ & $ {0.1656} $ & $ 0.3270 $ & $ 0.2645 $ & $ {0.4253} $ & $ {0.4114} $     \\
			&Deep-SQL	& $ 0.4141 $ & $ 0.3717 $ & $ 0.1030 $ & $ {0.1711} $ & $ 0.3330 $ & $ 0.2725 $  & $ {0.4336} $ & $ {0.4222} $    \\
			&MacridVAE  & $ 0.3828 $ & $ 0.3420 $ & $ 0.1031 $ & $ {0.1697} $ & $ 0.3002 $ & $ 0.2401 $  & $ {0.4050} $ & $ {0.3952} $    \\
			&Multi-VAE+ & $ 0.4128 $ & $ 0.3651 $ & $ 0.1076 $ & $ {0.1711} $ & $ 0.3385 $ & $ 0.2722 $  & $ {0.4386} $ & $ {0.4229} $    \\
			&DDCF	& $ \textbf{0.4304} $ & $ \textbf{0.3801} $ & $\textbf{ 0.1153} $ & $ \textbf{0.1819} $ & $ \textbf{0.3536} $ & $ \textbf{0.2847} $ & $ \textbf{0.4576} $ & $ \textbf{0.4415} $	\\
			\midrule
			\multirow{10}{2cm}{\emph{Amovie}}     & PMF  & $ {0.0098} $ & $ {0.0091} $ & ${ 0.0054} $ & $ {0.0091} $ & $ {0.0050} $ & $ {0.0033} $ & $ {0.0101} $ & $ {0.0099} $	\\
			&Primal-CR++	& $ 0.0247 $ & $ 0.0212 $ & $ 0.0077 $ & $ {0.0130} $ & $ 0.0143 $ & $ 0.0094 $ & $ {0.0262} $ & $ {0.0256} $       \\
			&DGCF  & $ 0.0683 $ & $ 0.0588 $ & $ 0.0287 $ & $ {0.0469} $ & $ 0.0420 $ & $ 0.0291 $  & $ {0.0734} $ & $ {0.0744} $      \\
			&SQL-Rank & $ 0.0719 $ & $ 0.0639 $ & $ 0.0289 $ & $ {0.0491} $ & $ 0.0452 $ & $ 0.0320 $ & $ {0.0784} $ & $ {0.0801} $       \\
			&MACR  & $ 0.0788 $ & $ 0.0690 $ & $ 0.0317 $ & $ {0.0543} $ & $ 0.0494 $ & $ 0.0346 $ & $ {0.0845} $ & $ {0.0860}$      \\
			&CPR  & $ 0.0827 $ & $ 0.0734 $ & $ 0.0362 $ & $ {0.0615} $ & $ 0.0520 $ & $ 0.0369 $ & $ {0.0897} $ & $ {0.0929} $     \\
			&DMF  & $ 0.0810 $ & $ 0.0699 $ & $ 0.0353 $ & $ {0.0581} $ & $ 0.0512 $ & $ 0.0354 $  & $ {0.0887} $ & $ {0.0891} $      \\
			&Deep-SQL	& $ 0.0851 $ & $ 0.0726 $ & $ 0.0374 $ & $ {0.0604} $ & $ 0.0549 $ & $ 0.0379 $  & $ {0.0891} $ & $ {0.0907} $      \\
			&MacridVAE  & $ 0.0923 $ & $ 0.0773 $ & $ 0.0412 $ & $ {0.0655} $ & $ 0.0608 $ & $ 0.0414 $  & $ {0.1017} $ & $ {0.1025} $      \\
			&Multi-VAE+ & $ {0.0916} $ & $ {0.0779} $ & $ {0.0404} $ & $ {0.0644} $ & $ {0.0594} $ & $ {0.0411} $  & $ {0.1002} $ & $ {0.1015} $      \\
			&DDCF	& $ \textbf{0.1008} $ & $ \textbf{0.0847} $ & $ \textbf{0.0456} $ & $ \textbf{0.0716} $ & $ \textbf{0.0655} $ & $ \textbf{0.0452} $ & $ \textbf{0.1106} $ & $ \textbf{0.1115} $	\\
			\midrule
			\multirow{10}{2cm}{\emph{Yahoo}}    & PMF  & $ 0.0608 $ & $ 0.0492 $ & $ 0.0816 $ & $ {0.1275} $ & $ 0.0360 $ & $ 0.0228 $  & $ {0.0809} $ & $ {0.0947} $     \\
			&Primal-CR++	& $ 0.0861 $ & $ 0.0668 $ & $ 0.1306 $ & $ {0.1945} $ & $ 0.0521 $ & $ 0.0324 $  & $ {0.1291} $ & $ {0.1493} $      \\
			&DGCF  & $ 0.1291 $ & $ 0.1008 $ & $ 0.1914 $ & $ {0.2852} $ & $ 0.0837 $ & $ 0.0534 $  & $ {0.1948} $ & $ {0.2239} $      \\
			&SQL-Rank & $ 0.1316 $ & $ 0.1012 $ & $ 0.2043 $ & $ {0.3040} $ & $ 0.0838 $ & $ 0.0524 $  & $ {0.2016} $ & $ {0.2333} $      \\
			&MACR  & $ 0.1358 $ & $ 0.1033 $ & $ 0.2252 $ & $ {0.3292} $ & $ 0.0887 $ & $ 0.0549 $ & $ {0.2193} $ & $ {0.2530}$      \\
			&CPR  & $ 0.1418 $ & $ 0.1101 $ & $ 0.2190 $ & $ {0.3254} $ & $ 0.0926 $ & $ 0.0587 $ & $ {0.2173} $ & $ {0.2517} $     \\
			&DMF  & $ 0.1450 $ & $ 0.1125 $ & $ 0.2227 $ & $ {0.3349} $ & $ 0.0946 $ & $ 0.0598 $  & $ {0.2246} $ & $ {0.2565} $      \\
			&Deep-SQL	& $ 0.1546 $ & $ 0.1187 $ & $ 0.2434 $ & $ {0.3574} $ & $ 0.1033 $ & $ 0.0650 $  & $ {0.2407} $ & $ {0.2773} $      \\
			&MacridVAE  & $ 0.1453 $ & $ 0.1107 $ & $ 0.2290 $ & $ {0.3332} $ & $ 0.0954 $ & $ 0.0597 $   & $ {0.2271} $ & $ {0.2603} $     \\
			&Multi-VAE+ & $ 0.1537 $ & $ 0.1161 $ & $ 0.2389 $ & $ {0.3442} $ & $ 0.1027 $ & $ 0.0643 $  & $ {0.2402} $ & $ {0.2728} $      \\
			&DDCF	& $ \textbf{0.1634} $ & $ \textbf{0.1234} $ & $ \textbf{0.2528} $ & $ \textbf{0.3649} $ & $ \textbf{0.1105} $ & $ \textbf{0.0695} $ & $ \textbf{0.2550} $ & $ \textbf{0.2884} $	\\
			\bottomrule
	\end{tabular}}
	\vspace{-2ex}
\end{table*}

\textbf{Baselines.} In the experiments, we compare our proposed approach with various stat-of-the-art explicit-feedback based methods, debiased CF methods, and disentangled CF methods:

\begin{itemize}
	\setlength{\itemsep}{0pt}
	\setlength{\parsep}{0pt}
	\setlength{\parskip}{0pt}
	\item \textbf{PMF:} Probabilistic Matrix Factorization~\citep{mnih2008probabilistic} is a classic pointwise rating prediction method.
	\item \textbf{Primal-CR++:} Primal-CR++~\citep{wu2017large} is a state-of-the-art pairwise CF approach for explicit feedback.
	\item \textbf{DGCF:} Disentangled Graph Collaborative Filtering is a state-of-the-art disentangled CF method for implicit feedback, which exploits user-item relationships through graphs.
	\item \textbf{DMF:} Deep Matrix Factorization~\citep{xue2017deep} is a state-of-the-art pointwise neural approach for explicit feedback.
	\item \textbf{MACR, CPR:} MACR~\citep{wei2021model} and CPR~\citep{wan2022cross} are two state-of-the-art debiasing approaches. LightGCN~\citep{he2020lightgcn} is used as the basic model for implementation of these two methods.
	\item \textbf{SQL-Rank, Deep-SQL:} Stochastic Queuing Listwise Ranking~\citep{wu2018sql} is a state-of-the-art listwise CF approach. Deep-SQL is the implementation for SQL-Rank with neural networks like DMF, which can produce better results.
	\item \textbf{MacridVAE:} MacridVAE is a state-of-the-art disentangled representation learning method based on user behaviors.
	\item \textbf{Multi-VAE+:} Variational Autoencoders for Collaborative Filtering (Multi-VAE)~\citep{liang2018variational} is a state-of-the-art autoencoder based approach. Here we build the model Multi-VAE+ by leveraging the normalized cross-entropy loss~\citep{xue2017deep} to replace the original cross-entropy loss for the reconstruction process so that Multi-VAE+ can handle explicit feedback. 
\end{itemize}

For all the above baselines, we used grid search to carefully tune the corresponding parameters, such as the number of dimensions and regularization parameters. Besides,
for SQL-Rank, we chose the ratio of subsampled unobserved items to positive items as $ 3:1 $ by grid search. For DMF, Deep-SQL, Multi-VAE+, DDCF-n, and DDCF, we use the same 2-layer MLP as the encoder architecture for a fair comparison. For MACR and CPR, we use the 3-layer LightGCN as the basic model. For our DDCF and DDCF-n models, we tuned the number of dimensions $ K $ and $ d $ in [50, 100, 150, 200, 250, 300] and the number of sampled intent channels $ L $ in [2, 3, ..., 10]. We set the prior parameter $ \alpha_k $ as 1 and tuned the penalty parameter $ \eta $ in $ [0.2,0.3,...,1.4,1.5] $. Moreover, to prevent posterior collapse, we also adopt the warm-up trick to increase $ \eta $ and decrease $ \tau $ by epochs. We set $ \tau=0.4, \tau_c = 0.2 $ and $ \lambda_2 =1, \lambda_3 =1,\lambda_4 =0.001 $. 

\vspace{-2ex}
\subsection{Overall Recommendation Performance}
We present the overall recommendation performance results for the three datasets in Table~\ref{tab2} under two types of settings, i.e., $ P = 5 $ and $ P = 10 $. We can discover from Table~\ref{tab2} that DDCF can outperform all the baseline methods on every dataset owing to the learned disentangled intent representations and preference representations. Specifically, DDCF outperforms the best baseline, by a relative boost of 5.85\%, 7.16\%, 4.46\%, 8.04\% for the metric P@$ 5 $, R@$ 5 $, MAP@5, and NDCG@5 in \emph{MovieLens}, 9.21\%, 10.68\%, 7.73\%, 8.75\% in \emph{Amovie}, and 5.69\%, 3.86\%, 6.97\%, 5.94\% in \emph{Yahoo}, respectively. Hence, the results clearly demonstrate the effectiveness of our proposed approaches. DGCF is designed for intent disentanglement in implicit feedback. Since they do not exploit the information of graded ratings, they cannot outperform the state-of-the-art recommenders for explicit feedback. Moreover, for MacridVAE, the preference and intent factors are still entangled in the rating modeling process. Therefore, DDCF can outperform MacridVAE. MACR and CPR are two debiased approaches. MACR is designed for popularity bias and CPR is for exposure bias. However, the two approaches are incapable of disentangling user intent and thus performs worse than DDCF. 
Besides, we can find that the matrix factorization based approaches, i.e., PMF, Primal-CR++, and SQL-rank, usually perform worse than deep learning based approaches. This observation demonstrates the effectiveness of neural networks. PMF and Primal-CR++ both perform not well in \emph{Amovie}. This may be because \emph{Amovie} is extremely sparse. Moreover,
we conduct the significant test (p-value = 0.05) to validate the improvements of DDCF over the strongest baseline are statistically significant in all three datasets.

\begin{table}[!t]
	\caption{Ablation experiments. }
	\vspace{-2ex}
	\label{tab:abl}
	\centering
	\resizebox{7.8cm}{!}{
		\begin{tabular}{llcccccc}
			\toprule
			Datasets & Methods  & P@10  & R@10 & MAP@10 & NDCG@10\\
			\midrule
			\multirow{3}{0.5cm}{\emph{MovieLens}} 
			&DDCF-n & $ 0.3680 $ & $ {0.1743} $ & $ {0.2736} $  & $ {0.4263} $    \\
			& DDCF-s  & $ 0.3754 $  & $ {0.1794} $ & $ 0.2802 $ & $ 0.4361 $     \\
			&DDCF	 & $ {0.3801} $ & $ {0.1819} $ & $ {0.2847} $  & $ {0.4415} $	\\
			\midrule
			\multirow{3}{0.5cm}{\emph{Amovie}} 
			&DDCF-n	 & $ {0.0802} $ & $ {0.0669} $  & $ {0.0421} $ & $ {0.1033} $       \\
			& DDCF-s    & $ 0.0826 $  & $ {0.0687} $ & $ 0.0443 $  & $ 0.1069 $    \\
			&DDCF  & $ {0.0847} $  & $ {0.0716} $  & $ {0.0452} $ & $ {0.1115} $	\\
			\midrule
			\multirow{3}{0.5cm}{\emph{Yahoo}} 
			&DDCF-n 	& $ 0.1187 $ &  $ {0.3510} $ & $ 0.0665 $ & $ 0.2797 $   \\
			& DDCF-s    & $ 0.1217 $  & $ {0.3593} $ & $ 0.0683 $ & $ 0.2857 $     \\
			&DDCF 	 & $ {0.1234} $ & $ {0.3649} $ & $ {0.0695} $ &  $ {0.2884} $	\\
			\bottomrule
	\end{tabular}}
	\vspace{-4ex}
\end{table}

\begin{figure*}[t!]
	\centering
	\includegraphics[width=0.45\textwidth]{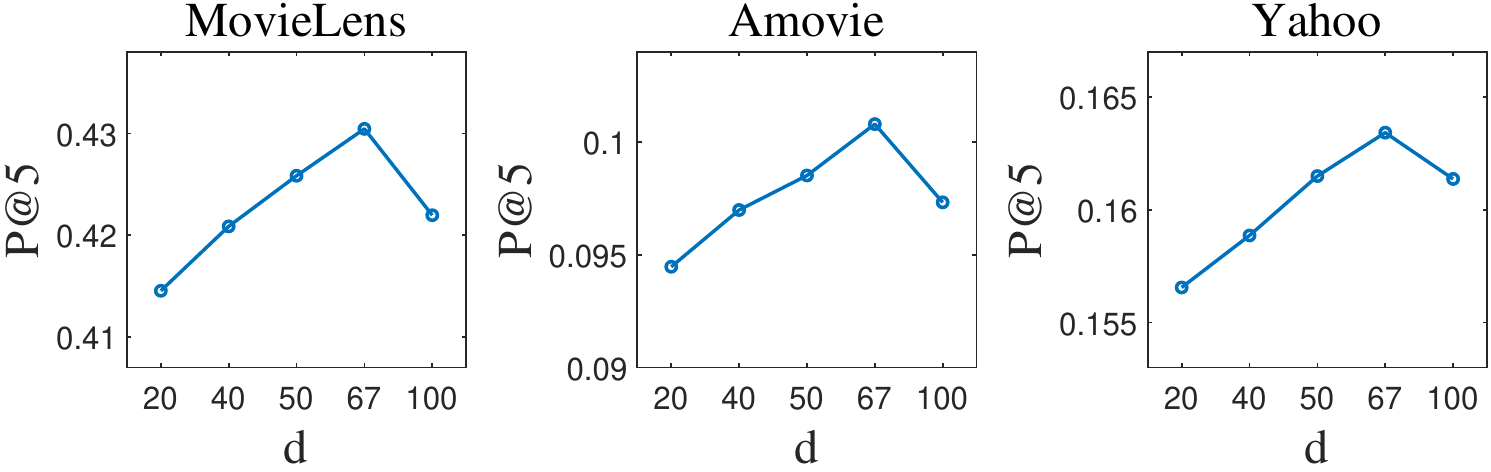}
	\vspace{-2ex}
	\caption{The performance of P@$ 5 $ with different values of dimension $ d $ on the three datasets.}
	\label{fig:rkind}
	\vspace{-2ex}
\end{figure*}

\begin{figure*}[!t]
	\centering
	\includegraphics[width=0.58\textwidth]{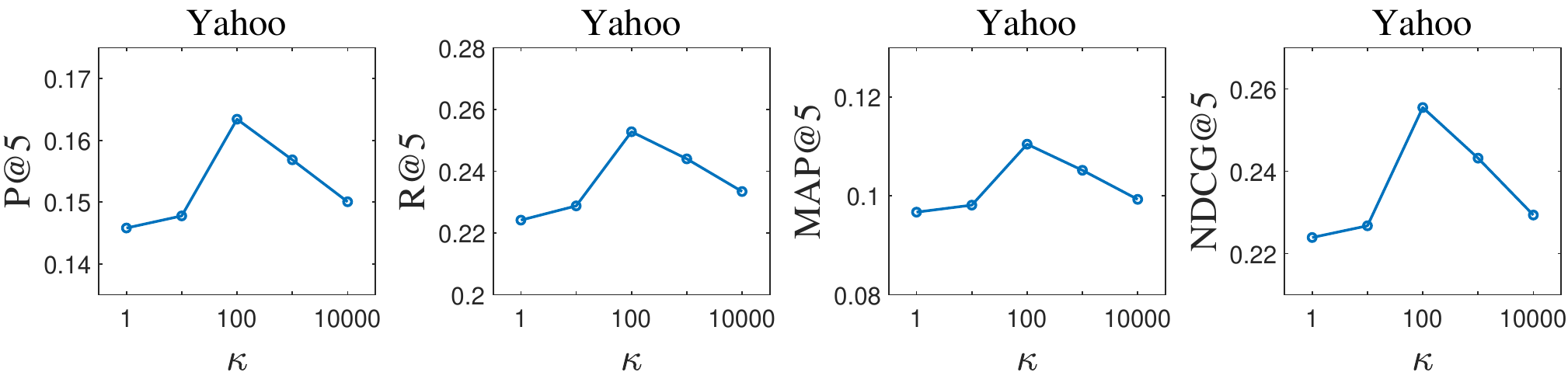}
	\vspace{-2ex}
	\caption{The performance of DDCF with different values of $ \kappa $ on the three metrics.}
	\label{fig:kkind}
	\vspace{-2ex}
\end{figure*}

\begin{table*}[t!]
	\caption{Case studies on user intents. (We present the top-3 intent channels for a user.)}
	\vspace{-2ex}
	\label{case}
	\centering
	\resizebox{15cm}{!}{
		\begin{tabular}{l|c}
			\toprule
			Intent Channel 1 & Payback, Scream 3 , Lethal Weapon 3 
			, Gone in 60 Seconds , Wild Wild West, Mission: Impossible 2   \\ 
			\cmidrule(r){1-2}
			Intent Channel 2  & A Bug's Life, Toy Story 2, Austin Powers: The Spy Who Shagged Me, Doctor Dolittle, Lethal Weapon          \\ 
			\cmidrule(r){1-2}
			Intent Channel 3	& Terminator 2: Judgment Day, The Matrix 
			, The Terminator  , Alien , The World Is Not Enough          \\   
			\bottomrule
	\end{tabular}}
	\vspace{-2ex}
\end{table*}

\vspace{-2ex}
\subsection{Ablation Study}
In this subsection, we evaluate two variants of DDCF. First, DDCF-n is the variant that only adopts positive feedback for intent recognition networks. Second, DDCF-s is the variant without disentangled contrastive learning mechanism. Comparing the results in \ref{tab:abl}, we can find that DDCF consistently outperforms DDCF-n in all the situations, which demonstrates again that low-rating items are not totally negative and can provide useful information from the intent perspective. With only positive feedbacks for intent modeling, DDCF-n cannot learn the intents well. As for DDCF-s, it always performs worse than DDCF, since it cannot leverage auxiliary supervision of positive pairs among different views. 
With contrastive learning, DDCF can learn the intention and preference better.
	
\vspace{-2ex}
\subsection{Investigations on Hyper-parameters}
In this paper, we factorize the rating matrix into the product of user and item latent representations under different intent channels in a low-rank space. Consequently, the number of dimensions $ d $ is quite vital for the performance. If $ d $ is too small, the latent space would have very weak representation ability to fit the real-world data. On the opposite, if $ d $ is too large, the model complexity would also become too large and may face the over-fitting problem.
In this subsection, we maintain the entire dimension $ Ld=200 $ to be unchanged and varied the number of sampled intents $ L $ and dimension $ d $ to train our model DDCF. The results in the evaluation metric precision are presented in Figure~\ref{fig:rkind} (see more in Appendix). We can observe that the performance result of DDCF is not good when $ r=10 $. With a larger value of $ d $, the performance tends to be much better. When $ d=67 $, DDCF achieves the best results. With larger $ d $, the performance of DDCF will begin to decrease.

Here, we also discuss the warm-up trick in our model. The posterior collapse problem will greatly influence the model training. When the KL divergence vanishes, the reconstruction results will be irrelevant to input records. This is unacceptable since we aim to provide personalized recommendations. To solve the problem, we use the warm-up trick for the KL divergences. Specifically, we add a penalty parameter to each KL term in the loss. Then, the initial value of the penalty parameter is set as 0. With the number of training batches growing, the penalty parameter will also increase linearly and finally reach the maximum value in the $ \kappa $-th batch. Here, we tune the value of $ \kappa $ for the loss in Equation 9 in the paper to control the influence of KL divergence. Then, we present the performances of P@5, R@5, and MAP@5 with $ \kappa=1, 10, 100, 1000, 10000 $ on the datasets \emph{Yahoo} in Figure~\ref{fig:kkind} as an example. We fix the maximum value of the penalty parameter as 1.
From Figure~\ref{fig:kkind} we can observe that when $ \kappa $ is small, the warm-up trick has almost no impact on the model training so that the performance of DDCF is quite bad. With the larger $ \kappa $, DDCF could greatly avoid the posterior collapse problem and achieve good results. However, when $ \kappa $ is too large, it will take too many batches to achieve the maximum value of the penalty parameter, which reduces the influence of KL divergence in the loss. Thus the model will converge into sub-optimal results. 

\subsection{Validation on Intent Channels}

In this subsection, we aim to verify the quality of the disentangled intent channels.
To validate the intent channels, one direct way is to compare our obtained intent channels with explicit item genres. Considering that the concepts in these two situations are not the same, the correlation would not be too strong. However, there should still exist some correlations, since item genre is one of the important reasons for users to interact with. Here we use the \emph{MovieLens-1m} dataset for validation, which contains 18 explicit genres of the movies. Since we cannot allocate which intent is related to which explicit genre, it is intractable to directly calculate the relevance between an intent and a genre. Instead, we choose to compute the successful co-occurrence rate for the item pairs. Specifically, we first count the number of pairs that exist in each of our generated intent channels. Then, if the pair also exists in any genre in the real data, we denote it as a successful co-occurrence pair. Hence, the successful co-occurrence rate can be defined as the ratio of the number of co-occurrence pairs to all pairs. Finally, we find that the successful co-occurrence rate is 47.6\% for our method DDCF in \emph{MovieLens-1m}. For comparison, the successful co-occurrence rate of MacridVAE is 38.8\% and the average successful co-occurrence rate for the random grouping setting is 33.0\%, which are significantly lower than DDCF. These results demonstrate that DDCF has automatically learned the information about movie genres from historical user-item interactions without the usage of ground-truth genre data.

\vspace{-2ex}
\subsection{Case Study}
In this subsection, we aim to provide interpretable insights into the learned intent channels. We set the number of intent channels $ K=200 $ and then trained DDCF on \emph{MovieLens}. Table~\ref{case} shows a real user case study, where we list the top-3 intent channels of one user. For each channel, we present the titles of the top movies. Thus, we can easily speculate about the user's interests with the help of the intent channels. First, it can be observed from Table~\ref{case} that Channel 1 and 3 are both comprised of action Movies. If the channels are coarse-grained, these two channels may be classified into one category. Thanks to our proposed independent sparse representations, DDCF can handle fine-grained intent channels. We can observe that movies in Channel 1 tend to contain more exciting and criminal elements, while movies in Channel 3 tend to be more scary and fictional.
Besides, the movies in Channel 2 are mainly animations and comedies, which shows another type of interests for this user. Consequently, DDCF can explicitly present the users' intents, which is beneficial for constructing user profiles. Moreover, conventional LFM tends to consider user preferences synthetically, and hence cannot distinguish the recommendation results. With DDCF, we can separately recommend movies with different intents according to users' dynamic requirements. For example, when the user wants to watch comedies, we can reduce the weights for Channel 1 and 3 while increase the weight for Channel 2 in the rating prediction process.

\vspace{-2ex}
\section{Conclusion}

In this paper, we proposed a two-fold representation learning approach, namely Double Disentangled Collaborative Filtering (DDCF), for improving the robustness and interpretability of recommender systems. A unique perspective of DDCF is that low-rating items can be partially used as positive feedback for recognizing intents, rather than always viewed as negative feedback in traditional approaches.
Specifically, the first-level disentanglement is for separating the influence factors of intent and preference, while the second-level disentanglement aims to construct independent sparse preference representations under different intents with limited computational complexity. Moreover, we designed a contrastive learning mechanism for disentangled representations. Finally, we conducted extensive experiments on three real-world datasets which validated both the effectiveness and interpretability of DDCF.

\bibliographystyle{ACM-Reference-Format}
\bibliography{main}

\end{document}